\documentclass[aps,superscriptaddress,nofootinbib12pt]{revtex4}
\usepackage{graphicx}
\usepackage{amssymb}
\usepackage{amsmath}
\usepackage[svgnames]{xcolor}
\usepackage{mathtools,slashed}
\usepackage[retainorgcmds]{IEEEtrantools}
\usepackage{physics}
\usepackage[compat=1.1.0]{tikz-feynhand}
\usepackage{tikz}
\usepackage{epstopdf}
\usepackage[utf8]{inputenc}
\usepackage{url}
\usepackage[colorlinks,citecolor=DarkGreen,linkcolor=DarkRed,urlcolor=DarkBlue]{hyperref}
\usepackage{float}

\usepackage{newtxtext}
\usepackage{newtxmath}
\usepackage{hyperref} 
\usepackage[makeroom]{cancel}
\usepackage{ulem}
\usepackage{epsfig}

\newcommand{\be}{\begin{eqnarray}}
\newcommand{\ee}{\end{eqnarray}}
\begin{document}

\title{Chiral vortical catalysis constrained by LQCD simulations}

\author{Rodrigo M. Nunes} \email{rodrigo.nunes@acad.ufsm.br }
\affiliation{Departamento de F\'{\i}sica, Universidade Federal de
  Santa Maria, Santa Maria, RS 97105-900, Brazil}  

\author{Ricardo L. S. Farias} \email{ricardo.farias@ufsm.br}
\affiliation{Departamento de F\'{\i}sica, Universidade Federal de
  Santa Maria, Santa Maria, RS 97105-900, Brazil}
  
\author{William R. Tavares}
\email{tavares.william@ce.uerj.br}
\affiliation{Departamento de Física Teórica, Universidade do Estado do Rio de Janeiro, 20550-013 Rio de Janeiro, RJ, Brazil}

\author{Varese S. Tim\'oteo} \email{varese@unicamp.br}
\affiliation{Grupo de \'Optica e Modelagem Num\'erica - GOMNI, Faculdade de Tecnologia - FT, Universidade Estadual de Campinas - UNICAMP, Limeira, 13484-332, SP, Brazil}


\begin{abstract}
Evidences of vortical effects have been recently found by experiments in heavy ion collisions, instigating new insights into the phase diagram of quantum chromodynamics (QCD). Considering the effect of rotations, lattice QCD data shows that the temperatures for deconfinement and chiral symmetry restoration should increase with real angular velocity, and the dominant effects are related to gluonic degrees of freedom. These findings could be essential for quark models in rotating systems that lack gluonic interactions, which predicts the decreasing of the chiral temperature transition with the angular velocity. To address this issue properly, in this work we apply the two-flavor Nambu--Jona-Lasinio model to explore the phase diagram in a rotating rigid cylinder with constant angular velocity in the mean field approximation. To circumvent the absence of gluons, we propose the application of an effective coupling dependent of the angular velocity, fitted to match the pseudocritical temperature of chiral phase transition in the model through lattice QCD data. Our results indicate that the running coupling induces the enhancement of the chiral condensate as a function of angular velocity, strengthening the breaking of chiral symmetry, an effect previously dubbed as chiral vortical catalysis. For the chiral susceptibility we observe stronger fluctuations around the transition temperature when we consider the running coupling. The phase diagram is affected by these findings shifting the critical end point (CEP) to higher temperatures and chemical potentials. 

\end{abstract}


\maketitle

\section{Introduction}

The extreme conditions achieved by the heavy ion collision (HIC) accelerators, e.g., RHIC and LHC, are enhancing our understanding about the phase structure of quantum chromodynamics (QCD). Strong experimental and numerical evidences indicate the existence of a crossover from a hadron gas to a quark-gluon plasma (QGP) at high temperatures and low baryon chemical potentials, as indicated by lattice QCD (LQCD) results \cite{Busza:2018rrf}, forming an almost perfect fluid \cite{Elfner:2022iae}.
Different situations and environments will potentially increment our understanding of strongly interacting quark matter, for instance, high values of baryon chemical potential and low temperatures, important to compact objects \cite{Fukushima:2010bq} and color superconductivity \cite{RevModPhys.80.1455}; strong magnetic and electric fields created by the initial states in peripheral heavy-ion collisions \cite{endrodi2024qcd}; and more recently, vorticity effects in HIC \cite{STAR:2017ckg}. In the latter, the STAR \cite{STAR:2022fan} and ALICE \cite{ALICE:2019aid} collaborations found evidences of vorticity associated with spin-alignment of $K^{*0}$ and $\phi$ vector mesons, but with different values of the spin density matrix element $\rho_{00}$, transforming our understanding of rotations in quark-gluon plasma into compelling, unanswered questions. However, it is possible to predict
 that noncentral collisions can generate a rapidly rotating QCD matter \cite{PhysRevLett.94.102301,PhysRevC.84.054910,Csernai:2013bqa,Jiang:2016woz,Deng:2016gyh,Becattini:2007sr}, with angular velocity of the order of $\omega\approx (9\pm 1)\times10^{21}$s$^{-1}$$\approx 7$ MeV \cite{STAR:2017ckg}. Hydrodynamics simulations, in turn, also anticipate rapidly rotating quark-gluon plasma (QGP) with angular velocities of $\omega\approx20-40$ MeV \cite{Jiang:2016woz}. Given the magnitude of the angular velocity, rotation effects can play a significant role in dynamics of QCD matter, as is the case of the prediction of vortex rings in event-by-event simulations \cite{DobrigkeitChinellato:2024xph}, the chiral vortical effect \cite{Kharzeev:2015znc,Huang:2020xyr} and the polarization of hyperons $\Lambda$ and $\bar{\Lambda}$ \cite{Ayala:2020soy,Ayala:2020jha,Xie:2019jun,Karpenko:2018erl,Yi:2023tgg,Kolomeitsev:2018svb,PhysRevD.96.096023}.

Recent developments in LQCD indicate that high angular velocities significantly influence the phase transition characteristics of QGP \cite{Braguta:2021ucr,Braguta:2021jgn,Braguta:2022str,Yang:2023vsw,Braguta:2023aio,Chernodub:2022veq}.
These studies reveal different behaviors for the deconfinement and chiral pseudocritical temperatures  as functions of the angular velocity when we consider rotations applied to quarks and gluons separately, i.e.,  they decrease in the former and increase in the latter.
 When both rotational regimes are incorporated, it results in a net elevation of the critical temperature. It is possible that the influence of gluons is dominant, since they are spin $1$ particles and its coupling with the orbital angular momentum could be more responsive to rotation than the spin $1/2$ particles \cite{Sun:2024anu}. The influence of gluons also proved to be decisive in determining the moment of inertia when considering rigid rotations, indicating the negative Barnet effect \cite{Braguta:2023tqz,Braguta:2023yjn,Braguta:2023qex}.

One way to test the LQCD results, avoiding the complication of highly computational efforts and extending our physical environment to baryonic densities is to apply effective approaches. The recent literature in this aspect focuses in applications in different versions of quark models, as the Nambu--Jona-Lasinio model (NJL) \cite{Jiang:2016wvv,Chernodub:2016kxh,Wang:2018sur,Sun:2023yux,Sun:2024anu,Sun:2021hxo,Ghalati:2023npr,Sadooghi:2021upd,Hua:2024bwn,Chen:2024utf,Sun:2023kuu,TabatabaeeMehr:2023tpt,Wei:2023pdf,Chen:2021aiq,Mehr:2022tfq,wei2022mass,Chen:2019tcp,Zhang:2018ome,Cao:2023olg}, 
 and linear-sigma model coupled with quarks \cite{Singha:2024tpo,Hernandez:2024nev, Chernodub:2024wis,Chen:2023cjt,Wan:2020ffv,Chen:2020ath}; as well as holographic approaches \cite{Chen:2024jet,Braga:2023qej,Braga:2023qee,Zhao:2022uxc,Yadav:2022qcl}, scalar theories \cite{siri2024bose,Siri:2024scq,Ambrus:2023bid}, hadron resonance gas model \cite{Pradhan:2023rvf,Fujimoto:2021xix} and effective chiral approaches \cite{Chen:2024tkr,Gaspar:2023nqk}. Most effective quark models predict a different behavior from LQCD, wherein rotation expedites the phase transition, thereby reducing the critical temperature \cite{Jiang:2016wvv,Chernodub:2016kxh,Sun:2021hxo,Sun:2023yux,Wang:2018sur,Singha:2024tpo,Hernandez:2024nev}. 
In particular, different attempts to describe QCD matter under rotation in the NJL model have been explored, since it originally does not include gluonic degrees of freedom, which is essential to reproduce LQCD results \cite{Braguta:2021jgn,Braguta:2022str}. {Usually it is possible to include the Polyakov loop potential to extend the NJL model (PNJL) in order to obtain the confinement/deconfinement transition \cite{Meisinger:2001cq,Fukushima:2003fw,Ratti:2005jh,Fukushima:2008wg}. A reparametrization of the Polyakov loop potential has been proposed in order to make it as a function of the angular velocity, which turns the deconfinement and chiral pseudocritial temperatures increasing functions of angular velocity \cite{Sun:2024anu}. An attempt to mimic gluons in the NJL model was made in Ref. \cite{Jiang:2021izj}, in which the author established  the concept of chiral vortical catalysis, in which it is introduced a coupling that depends linearly on the angular velocity, inducing an increasing of the critical temperature and slowing down the melting of the condensate with the temperature.  Other applications to investigate the confinement/deconfinement transition with rotation can be found in the following references \cite{Chernodub:2020qah,Singha:2024tpo,Chen:2020ath,Mameda:2023sst,Fujimoto:2021xix,Zhao:2022uxc,Jiang:2023zzu,Jiang:2024zsw}.

The disagreement between some predictions of LQCD and effective model results are a well-known subject, indicating that some adjustments have to be explored in order to match both approaches. For instance, in the context of magnetic fields, LQCD predicts the inverse magnetic catalysis, in which the pseudocritical temperature of deconfinement and chiral phase transitions decreases as function of the magnetic field, accompanied by a nonmonotonic behavior of the chiral condensate for $eB\gtrsim 0.2$ GeV$^2$, which is not reproduced by effective model methods in the mean field approximation \cite{RevModPhys.88.025001}.
 A common approach to match such discrepancies is to implement a magnetic or thermo-magnetic dependence of the coupling constant of the model fitted by LQCD data \cite{Tavares:2021fik,Farias:2016gmy,Endrodi:2019whh,Moreira:2021ety,PhysRevC.90.025203,PhysRevD.102.014032,Avancini:2016fgq,Ferreira:2014kpa}.
 Inspired by these findings, in this work we use the two-flavor NJL model in a rigid, finite-size cylinder rotating with a constant angular velocity with the implementation of a running coupling as a function of the angular velocity, fitted by results of the pseudocritical temperature of chiral transition obtained by LQCD \cite{Braguta:2022str}. This is a more straightforward way to include the vortical effects induced by gluonic degrees of freedom on the model, without leading with more complex parametrizations of the Polyakov loop \cite
{Ratti:2005jh} or beyond mean field approaches and its nontrivial parametrizations \cite{zhuang1994thermodynamics}.  Within this schematic model structure, we explore basic aspects of the model as the chiral condensate, the effective quark masses, the chiral susceptibilities, and the phase diagram of the quark model. All of the present results are compared with and without the running couplings.

This work is structured as follows: in Sec. \ref{sec2} we describe the formalism, and how the running coupling is determined. Section \ref{sec3} we present our results and discuss the effective quark mass, chiral condensate, chiral susceptibility and the phase diagram. In Sec. \ref{sec4} we show a summary and the conclusions of our work.

\section{Thermodynamic potential of the two-flavor NJL model with rotation}\label{sec2}

In order to introduce vortical effects in the two-flavor NJL model we use the same approach previously developed in the literature  \cite{Chernodub:2016kxh,Jiang:2016wvv,Wang:2018sur}. Then, we consider a rigid cylinder rotating around the fixed $\hat{z}$ axis with constant angular velocity $\Vec{\omega} = \omega\hat{z}$ , in which the space-time metric, $g_{\mu\nu}$, is given by

\begin{eqnarray}
g_{\mu\nu}=\left(
\begin{array}{cccc}
 1-(x^2+y^2)\omega^2 & y\omega & -x\omega& 0 \\
 y\omega & -1 & 0 & 0 \\
 -x\omega & 0 & -1 & 0 \\
 0 & 0 & 0 & -1 \\
\end{array}
\right).
\label{tensor}
\end{eqnarray}

We assume in this system that we have a rigid rotation, i.e., that all spatial regions of the fermionic system rotates with the same angular velocity $\omega$. In this way the system is limited in the perpendicular direction of the $\hat{z}_3$ axis, which we can infer that the absolute velocity of a point in a distance $r$ of the axis should always be $r\omega\leq 1$, to obey causality \cite{Chernodub:2016kxh}. Moreover, in the present work we will treat the chiral condensates as homogeneous, i.e., they do not depend on the coordinate system. This is a frequently applied approximation that is valid for moderately small angular velocities \cite{Chernodub:2016kxh,Chen:2021aiq,Singha:2024tpo} and enables us to compare our results with the similar literature. For rapid rotating systems, notable radial inhomogeneities can be important, as indicated by Refs. \cite{Chernodub:2020qah,Chernodub:2022veq,Braguta:2023iyx}.

In this prescription the lagrangian of the two-flavor NJL model  assumes the following form,

\begin{equation}
   \mathcal{L}_{\text{NJL}} =\bar{\psi} \biggl(   {i\bar{\gamma} ^\mu (\partial _\mu+\Gamma_{\mu})- \hat{m}+\gamma ^0 \mu } \biggr)\psi + G\left[{\left( {\bar \psi \psi } \right)^2}+\left( {\bar \psi i\gamma_5\Vec{\tau} \psi } \right)^2\right],\label{LNJL}
\end{equation}
where $\hat{m}=\text{diag}(m_u \quad m_d)$ is the quark mass matrix, in which we consider $m_u=m_d=m$; $\psi$ is the spinor representing the quark fields, $\psi=(\psi_u \quad\psi_d)^{T}$; $G$ is the coupling constant; $\mu$ is the chemical potential; and $\vec{\tau}$ represents the Pauli Matrices. The gamma matrices in curved space are given by $\bar{\gamma}^\mu= e^{\mu}_i\gamma^i$, with $\gamma^i$ being the usual gamma matrices and $e^{\mu}_i$ the  tetrads for spinors. The connection $\Gamma_{\mu}$ is defined as,
\begin{eqnarray}
    \Gamma_{\mu}  = -\frac{i}{4}\omega_{\mu ij}\sigma^{ij},\hspace{1cm}\omega_{\mu ij} =g_{\alpha\beta}e^{\alpha}_i(\partial_{\mu}e^{\beta}_j+\Gamma^{\beta}_{\mu\nu}e^{\nu}_j), \hspace{1cm} \sigma^{ij} = \frac{i}{2}\left[\gamma^i,\gamma^j\right],
\end{eqnarray}
where $\Gamma^{\beta}_{\mu\nu}$ is the Christoffel connection associated to $g_{\mu\nu}$ \cite{Chernodub:2016kxh}. In this work we adopt the following definition for the tetrads in a constant rotating system around the $\hat{z_3}$ axis \cite{Chen:2015hfc},

\begin{eqnarray}
    e^t_0=e^x_1=e^y_2=e^z_3=1, \hspace{1cm} e^x_0=\omega y, \hspace{1cm} e^y_0=-\omega x,
\end{eqnarray}
in which the remaining components are zero. With these definitions and using the mean field approximation, equation (\ref{LNJL}) is rewritten as \cite{Sun:2021hxo,Jiang:2016wvv},

\begin{eqnarray}
    \mathcal{L}_{NJL} = \bar{\psi}\left[i\gamma^{\mu}\partial_{\mu}-M+\gamma^0\mu+\gamma^0\left((\Vec{\omega}\times\Vec{x})\cdot(-i\Vec{\partial})+\Vec{\omega}\cdot \Vec{S}_{4\times 4}\right) \right]\psi-\frac{(M-m)^2}{4G},
\end{eqnarray}
where $M = m-2G\sum_{f=u}^{d}\langle \bar{\psi_f}\psi_f\rangle$, is the gap equation, which determines the effective quark mass $M$ self-consistently;  $\langle \bar{\psi_f}\psi_f\rangle$ is the chiral condensate for each flavor and $\Vec{S}_{4\times 4}=\frac{1}{2}\text{diag}(\Vec{\sigma},\Vec{\sigma})$. We can identify $\omega J^z = (\Vec{\omega}\times\Vec{x})\cdot(-i\Vec{\partial})+\Vec{\omega}\cdot \Vec{S}_{4\times 4}$ with $J^z$ the $z$-component of total angular momentum which the eigenvalues are given by $\left(n+\frac{1}{2}\right)$. The inverse quark propagator is \cite{Sun:2021hxo},
\begin{eqnarray}
    D^{-1} = \beta\left[\gamma^0\left(-i\omega_N+\mu+\left(n+\frac{1}{2}\right)\omega\right)-\left(\gamma^kp_k+M\right)\right],
\end{eqnarray}

\noindent in which the imaginary time formalism was used, with $\omega_N=(2N+1)\pi T$ being the Matsubara frequencies and $\beta$  the inverse temperature, $1/T$. Applying standard finite temperature quantum field theory technics one can derive the thermodynamic potential of the two-flavor NJL model for rotating fermions at finite temperature and quark chemical potential,

\begin{eqnarray}
    \Omega_{NJL} &=& \frac{(M-m)^2}{4G} - \frac{3}{2\pi^2}\sum_{n=-\infty}^{\infty}\int_0^{\Lambda}p_tdp_t\int_{-\sqrt{\Lambda^2-p_t^2}}^{\sqrt{\Lambda^2-p_t^2}}dp_z\mathcal{J}_n(p_t,r)\epsilon_n\nonumber\\
    &-& \frac{3}{2\pi^2}\sum_{n=-\infty}^{\infty}\int_0^{\infty}p_tdp_t\int_{-\infty}^{\infty}dp_z\mathcal{J}_n(p_t,r)T\biggl[\ln\biggl(e^{-\frac{\epsilon_n+\mu}{T}}+1\biggr) +\ln\biggl(e^{-\frac{\epsilon_n-\mu}{T}}+1\biggr)\biggr],\label{OmegaNJL}
\end{eqnarray}
where we define $\mathcal{J}_n(p_t,r)=\biggl(J_{n+1}(p_tr)^2+J_{n}(p_tr)^2\biggr)$ with $J_n(x)$ being the Bessel function of the first kind and $\epsilon_n = \sqrt{M^2+p_z^2+p_t^2}-(\frac{1}{2}+n)\omega$ is the quark energy dispersion. Since the NJL model is a nonrenormalizable theory in $3+1$ D, we choose the sharp cutoff scheme as the regularization procedure, which introduces the cutoff $\Lambda$ as a new parameter. The parametrization of the model will be given in the numerical results section, Sec. \ref{sec3}. 

For completeness, we can obtain the explicit result for the gap equation from the minimization of the thermodynamic potential, i.e., $\partial \Omega/\partial M =0$,

\begin{eqnarray}
    && \frac{M-m_0}{2G}-\frac{3}{2\pi^2}\sum_{n=-\infty}^{\infty}\int_{0}^{\Lambda}p_tdp_t\int_{-\sqrt{\Lambda^2-p_t^2}}^{\sqrt{\Lambda^2-p_t^2}}dp_z \mathcal{J}_n(p_t,r)\frac{M}{\sqrt{M^2+p_t^2+p_z^2}}\nonumber\\
    &+&\frac{3}{2\pi^2}\sum_{n=-\infty}^{\infty}\int_{0}^{\infty} p_tdp_t\int_{-\infty}^{\infty}dp_z \mathcal{J}_n(p_t,r) \frac{M}{\sqrt{M^2+p_t^2+p_z^2}}\biggl[\frac{1}{e^{\frac{\epsilon_n-\mu}{T}}+1}+\frac{1}{e^{\frac{\epsilon_n+\mu}{T}}+1}\biggr]=0.\label{Gapnjl}
\end{eqnarray}



The effect of rotating gluons can be introduced via running coupling $G(\omega)$ \cite{Jiang:2021izj}, in which we use lattice data as an input. For this purpose we start by using the same relation of the LQCD for the pseudocritical temperature as a function of the angular velocity \cite{Braguta:2022str}, i.e.,
\begin{eqnarray}
    \frac{T_c(v)}{T_c(0)} = 1+B_2\frac{v^2}{c^2},\label{TcLattice}
\end{eqnarray}
where $v=\omega \frac{L_s}{2}$, $L_s$ is the lattice spacing; $B_2$ depends on the lattice parameters as well as the ratio of pseudo scalar to vector meson masses, which in this work we choose $m_{ps}/m_v=0.80$ so $B_2\approx 1.13153$. 
 In the next step, we used the lattice QCD pseudocritical temperature $T_c$ at several values of the angular velocity $\omega$, and determined the value of the NJL coupling required to reproduce the lattice QCD calculation of $T_c$ at each value of the angular velocity.

In addition to the effects of rotation in the chiral condensate and phase diagram, we are also interested in investigating the chiral susceptibility, defined as 


\begin{eqnarray}
    \chi&=&\frac{1}{m_{\pi}^2}\frac{\partial \langle \bar{\psi}\psi \rangle}{\partial T},
\end{eqnarray}
in which, $m_{\pi}=141.715$ MeV, is the pion mass value in the vacuum.



\section{Numerical Results}\label{sec3}

To evaluate the numerical results, in this work we adopt the sharp cutoff regularization scheme, with the following set of  parameters: the cutoff  $\Lambda = 651$ MeV; the bare quark mass $m = 5.5$ MeV, and the coupling constant $G=5.04$ GeV$^{-2}$ \cite{Wang:2018sur}. These values are chosen in order to obtain the vacuum values of the pion mass $m_{\pi}=141.715$ MeV, the pion decay constant $f_{\pi}=94.043$ MeV, and the chiral condensate $\langle \overline{\psi_u}\psi_u \rangle^{1/3}=-(251.322)$ MeV.
In order to avoid causality violation, the condition $\omega r< 1$ needs to be ensured, therefore we choose $r=5$ GeV$^{-1}$ which means that the angular velocity can assume the maximum value at $\omega=0.2$ GeV.
 Following the procedure described  at the end o Sec.\ref{sec2}, the running coupling is chosen to be
\begin{eqnarray}
    G(\omega) = G_\alpha + G_\beta ~ \exp(\frac{\omega}{\Omega} )\;,\label{Gw}
\end{eqnarray}
 where $G_\alpha= 4.97667$ GeV$^{-2}$, $G_\beta = 0.05840$ GeV$^{-2}$, $\Omega=0.02457$ GeV and $\omega$ is the angular velocity. These parameters are determined by fitting the couplings, at different velocities, to the shifted exponential in Eq.(\ref{Gw}), and correspond to the minimum of the root mean square error. In the range of angular velocities considered, with three parameters, we achieved a very good agreement between $T_c \times \omega$ and the lattice results~\cite{Braguta:2022str}. Also, the mathematical expression given in Eq.(\ref{Gw}) is particularly chosen since the plot for $G\times\omega$ resembles a shifted exponential.


In Fig. \ref{condxT}, we show the chiral condensate as a function of the temperature at vanishing chemical potential and different values of $\omega$. The results are shown for constant coupling, in the left panel, and running coupling, in the right panel. Considering the constant coupling we have obtained consistent results when compared with the previous literature findings with effective model approaches, in which the partial restoration of chiral symmetry is enhanced by increasing $\omega$ and $T$ \cite{Jiang:2016wvv,Sun:2021hxo,Chernodub:2016kxh}. Rather, we observe  that vorticity has negligible impact in the low-temperature regime in accordance with previous findings from Ref. \cite{Chernodub:2016kxh}. When introducing an $\omega$-dependent coupling, a phenomena that can be referred to as chiral vortical catalysis \cite{Jiang:2021izj} occurs, where the critical temperature increases with increasing angular velocity. Also, the chiral condensate is catalyzed at low temperatures, which is better illustrated in Fig. \ref{condxomega}.


\begin{figure}[!ht]
	\begin{center}
		\includegraphics[scale= 0.42]{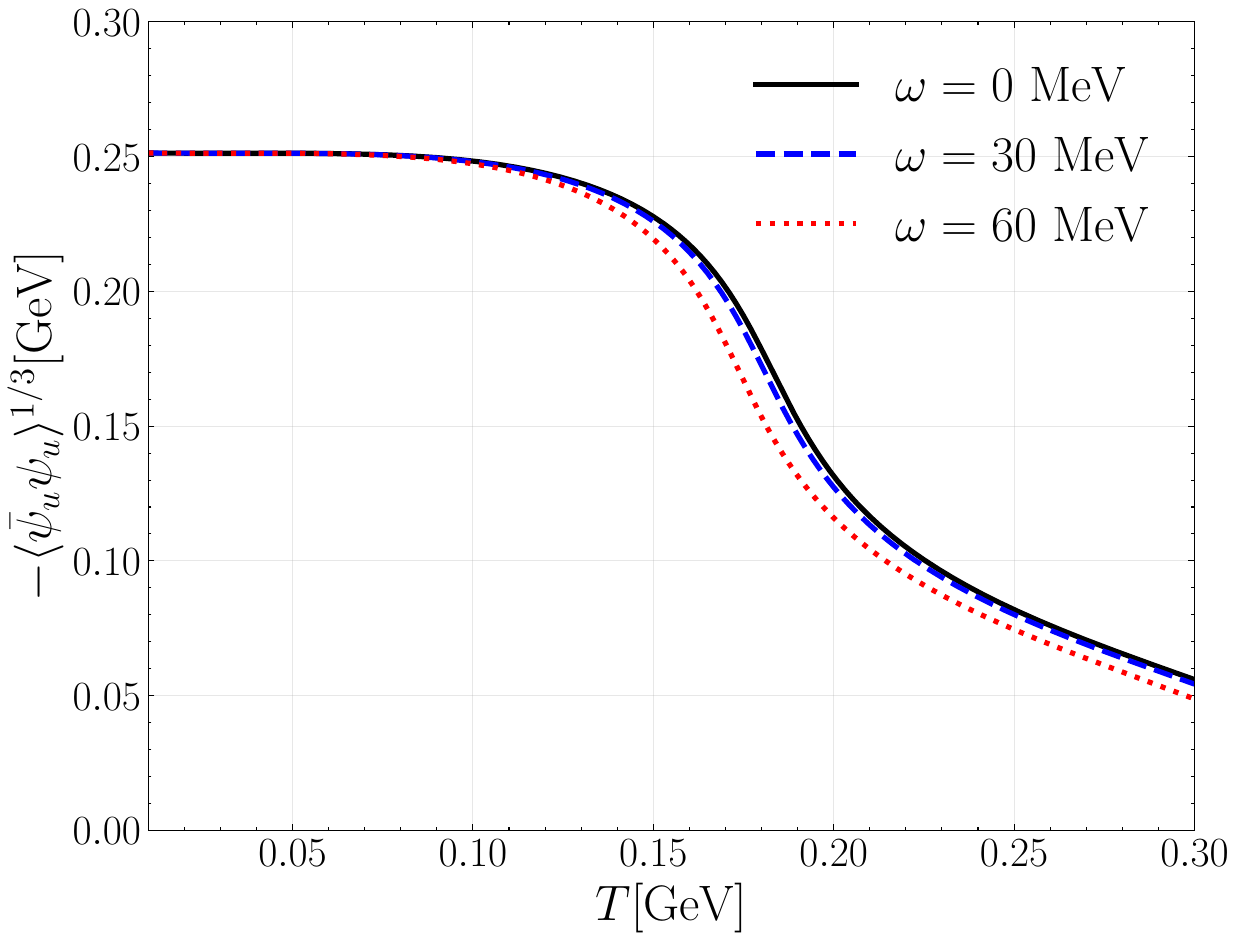}
            \includegraphics[scale= 0.42]{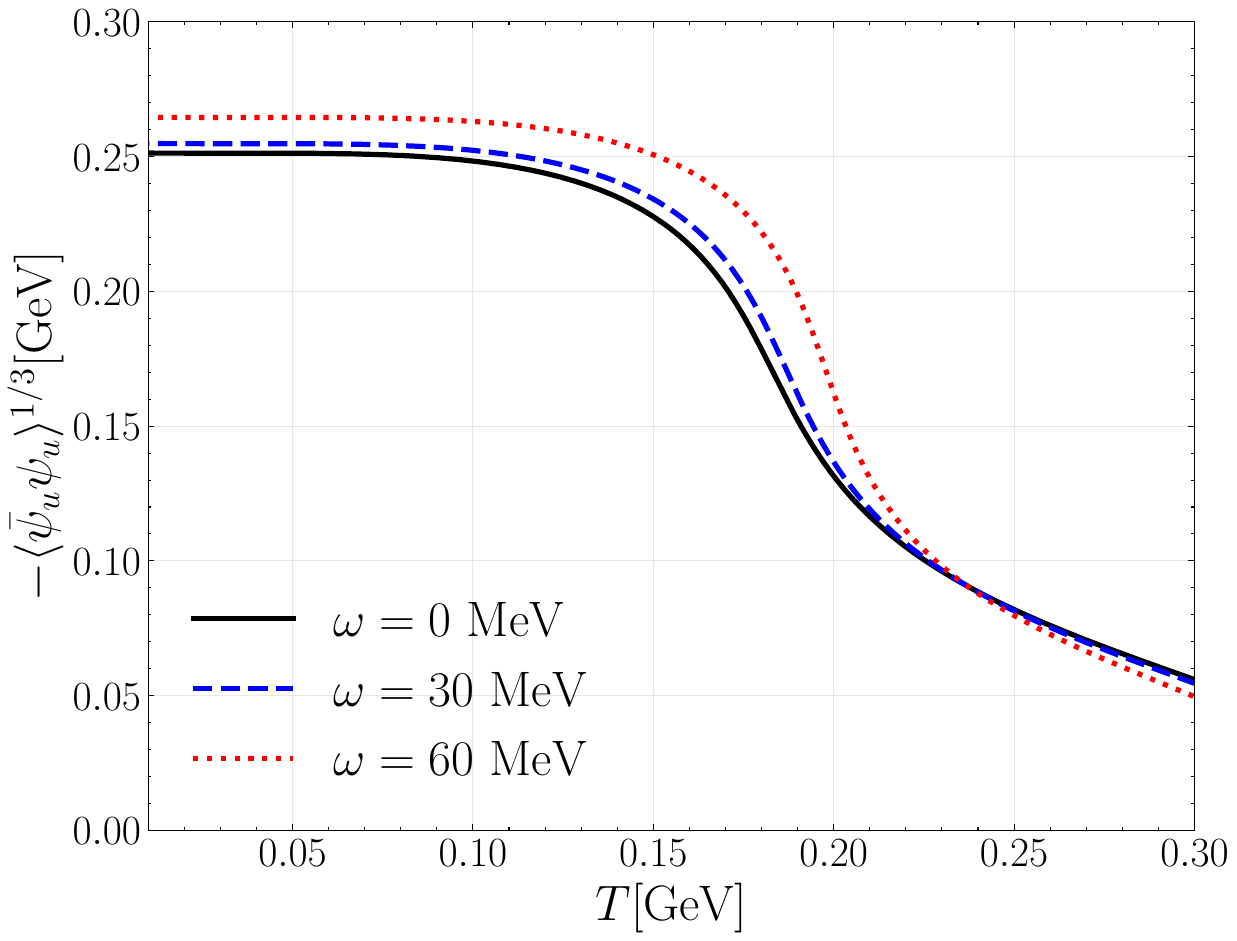}
	\end{center}
	\caption{Chiral condensate as a function of the temperature with fixed value of angular velocity. In the left panel, the case with fixed $G$, and in the right panel, the case with $G(\omega)$.}\label{condxT}
\end{figure} 

Figure \ref{condxomega} shows the chiral condensate as a function of the angular velocity at different temperature values for constant (left panel) and running (right panel)  couplings. With the fixed coupling at $T=50$ MeV, the chiral condensate is only significantly affected after $\omega\gtrsim 0.15$ GeV, when the angular velocity suppresses its values. For temperature $T=150$ MeV the effect of $\omega$ becomes more relevant, with the suppression of the condensate starting at $\omega\sim 0.05$ MeV, showing that the rotation in the system is more relevant to high temperature systems, which is confirmed when we look to the effects at $T=200$ MeV. On the other hand, when we look to the results with the running coupling at fixed temperatures, we observe the increase of the chiral condensate for the whole range of angular velocities considered, which is a direct consequence of the coupling being a function that increases with angular velocity.  Hence, with the adjustment in the coupling we can completely change the behavior of the chiral condensate, which is now enhanced by the angular velocity inducing a stronger breaking of the chiral symmetry as predicted by the chiral vortical catalysis \cite{Jiang:2021izj}, in accordance with  the predictions of Ref.  \cite{Yang:2023vsw}. This is, therefore, a strong evidence that gluonic degrees of freedom are indeed crucial missing pieces in low energy effective approaches, as the NJL model, when considering rotations in the system. It is important to note that we have different ranges of $\omega$ in both plots of Fig. \ref{condxomega}, which is justified by the maximum value of $\omega r \sim 0.3$ adopted by LQCD \cite{Braguta:2022str}, while our fitting function of $G(\omega)$ is valid also as an extrapolation of this limit, i.e., it is applied until we reach $\omega r \sim 0.55$. Since our parametrization considers $r=5$ GeV$^{-1}$, we find the range of the right panel.

\begin{figure}[!ht]
	\begin{center}
        \includegraphics[scale= 0.42]{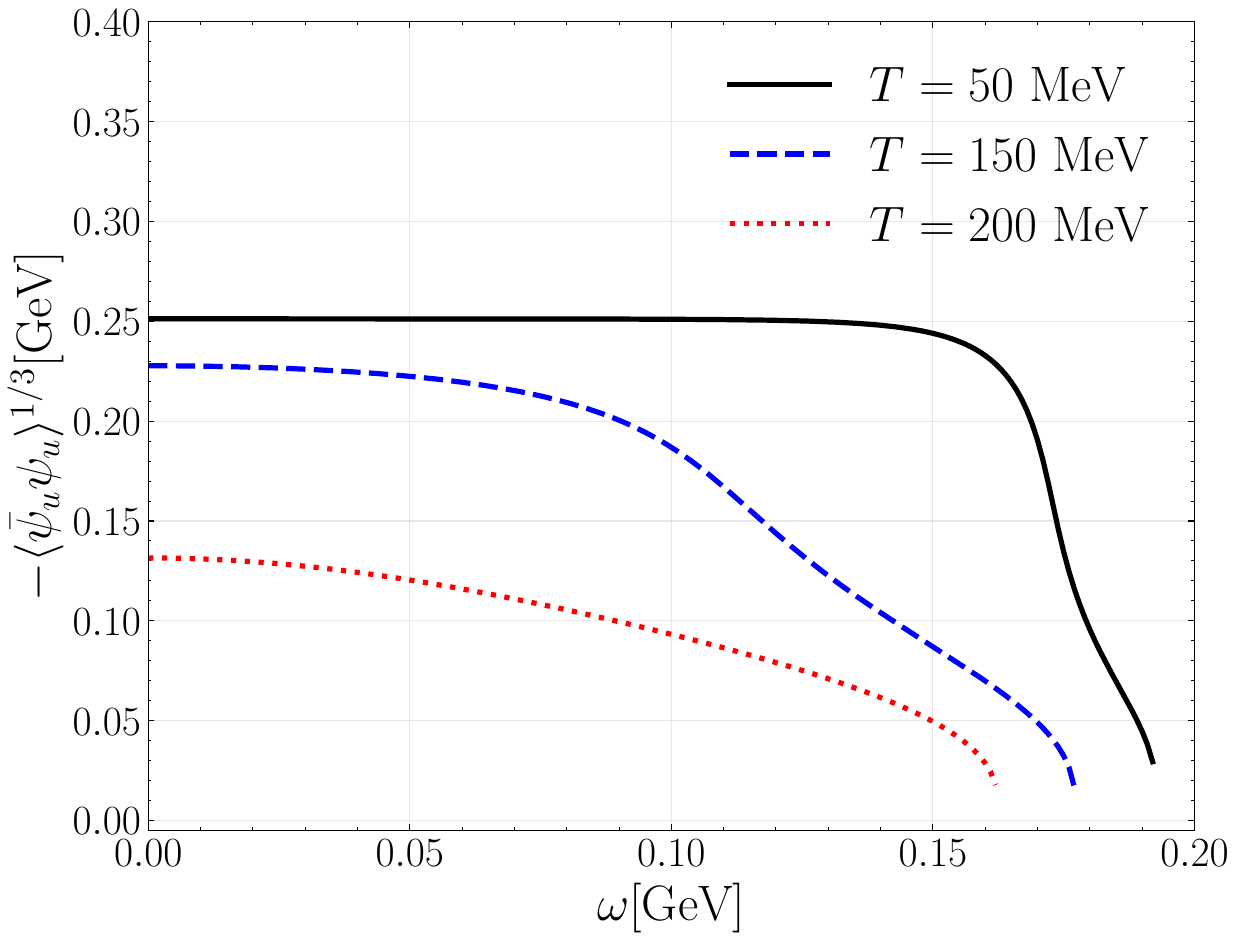}
        \includegraphics[scale= 0.42]{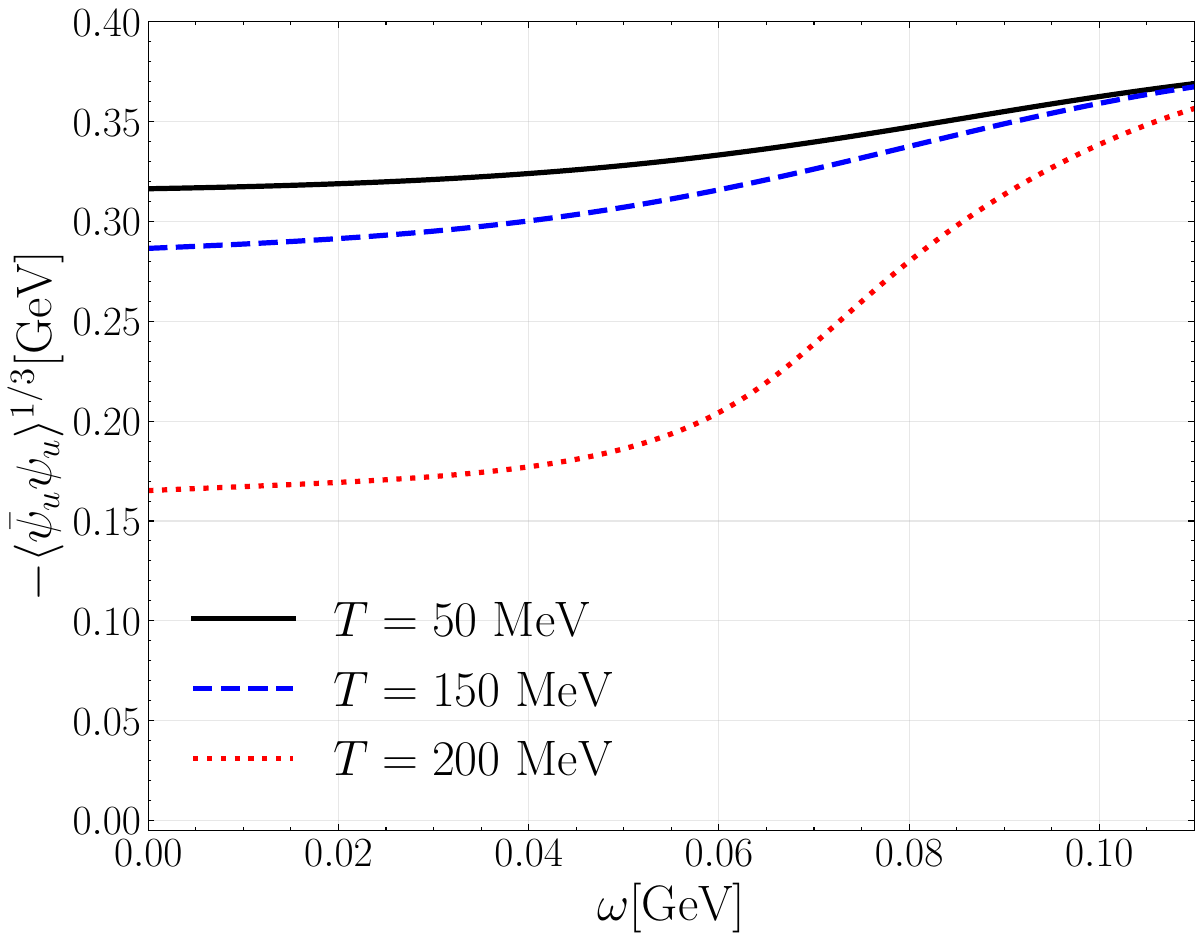}
	\end{center}
	\caption{Chiral condensate as a function of the angular velocity and fixed values of temperature. In the left panel we consider the case with fixed $G$, while in the right panel the case with $G(\omega)$.}\label{condxomega}
\end{figure}

For completeness we also analyse the effective quark masses as a function of the temperature for different values of angular velocity in Fig.\ref{MxT}. Since the effective quark masses are proportional to the chiral condensates, i.e., $M\propto G\sum_{f=u}^d\langle \bar{\psi}_f\psi_f\rangle$, it is clear that for fixed coupling, $G$, the behavior of the quark masses is qualitative the same as observed in left panel of Fig. \ref{condxT}. However, the running coupling induces the effective quark masses to be enhanced at low temperatures as we increase the angular velocity. This effect is more clear in the case of $\omega=60$ MeV, showing that the quark masses are more affected than the condensates, increasing $22\%$ when compared to $\omega=30$ MeV case, as observed in right panel of Fig. \ref{MxT}. For comparison, the quark condensate increases $12\%$ in the same values of $\omega$ and $T$. For the high temperature limit, the masses with different values of $\omega$ almost coincide. In order complete the analysis, we have also calculated the effective quark masses as a function of the angular velocities, and the results follows the same qualitative behavior predicted observed in Fig. \ref{condxomega}, then, for simplicity, we omit such plot.

\begin{figure}[!ht]
	\begin{center}
        \includegraphics[scale= 0.42]{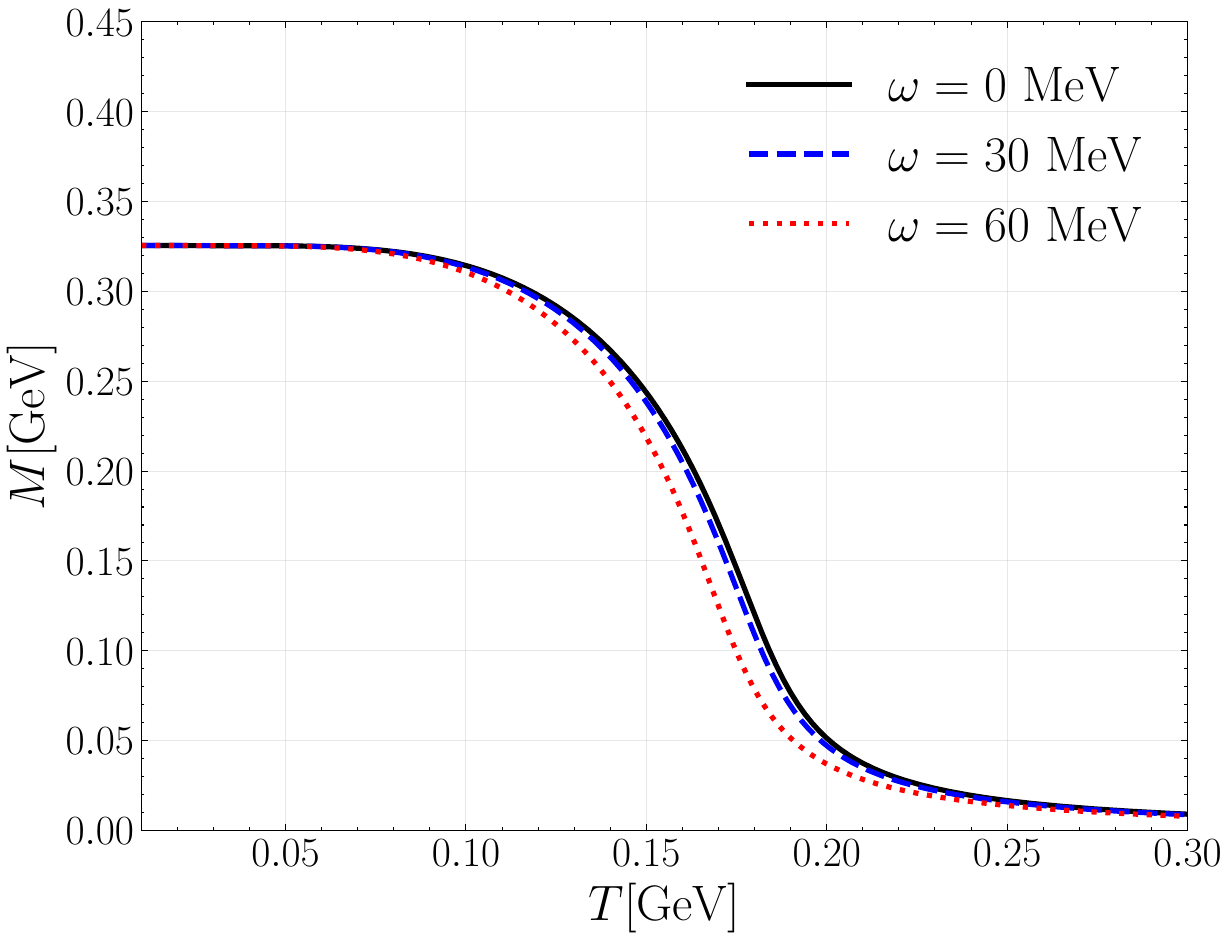}
        \includegraphics[scale= 0.42]{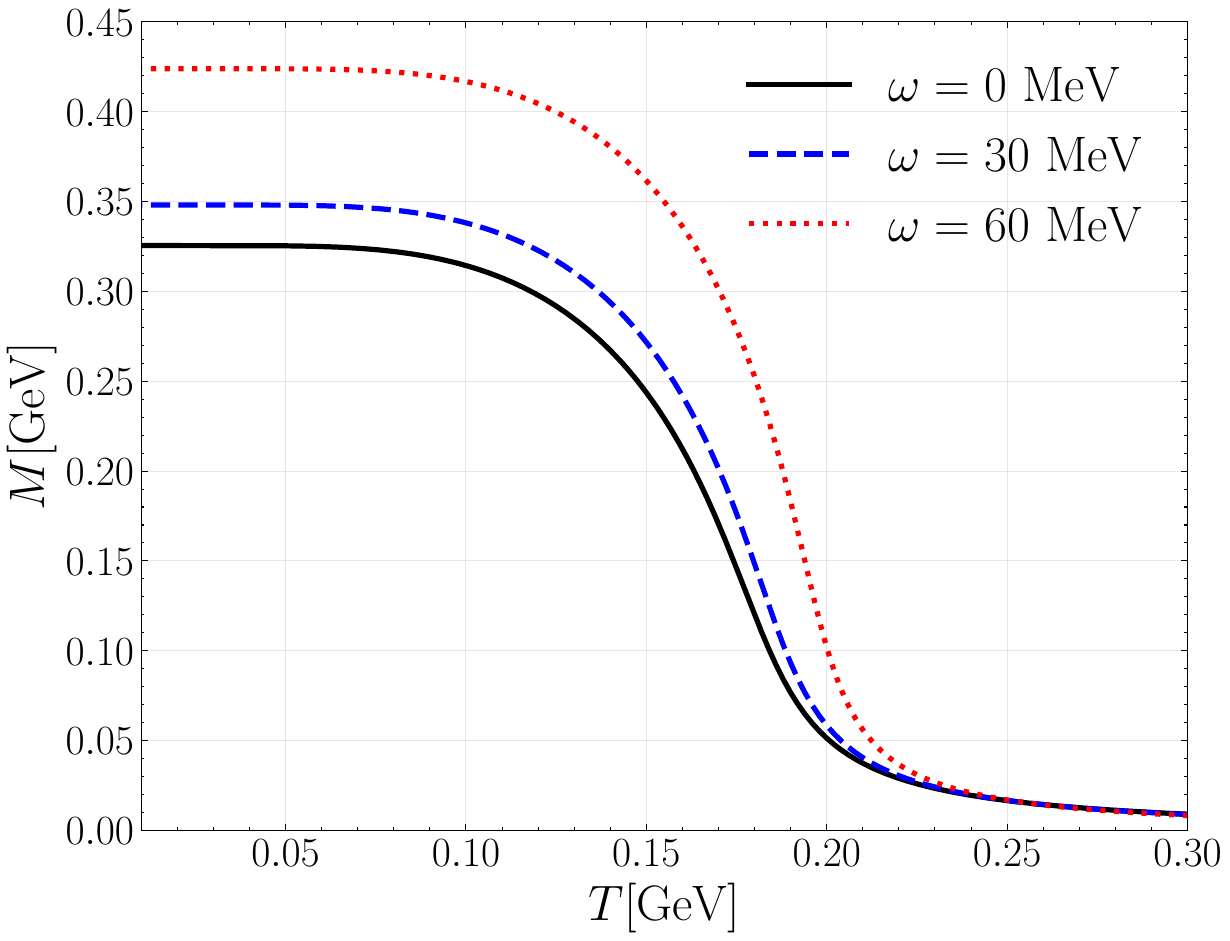}
	\end{center}
	\caption{Effective quark masses as a function of the angular velocity and fixed values of temperature. In the left panel we consider fixed $G$, while in the right panel $G(\omega)$.}\label{MxT}
\end{figure}

In order to study the structure of the pseudocritical temperature of chiral phase transition we show in Fig.\ref{dcondxT} the chiral susceptibility as a function of the temperature for different values of angular velocity, with both couplings. In both plots the peaks indicate the crossover temperature transition. In the left plot, with constant coupling, the peak shifts to lower temperatures with a slightly enhancement in its magnitude as we increase the angular velocity. On the other hand, when we consider the running coupling, the peak shifts to higher temperatures with the angular velocity, showing a noticeable enhancement on its magnitude. These results show that the effects of rotations guided by our effective gluonic interaction changes how the system is affected by fluctuations close to the transition temperature.

\begin{figure}[!ht]
	\begin{center}
        \includegraphics[scale= 0.42]{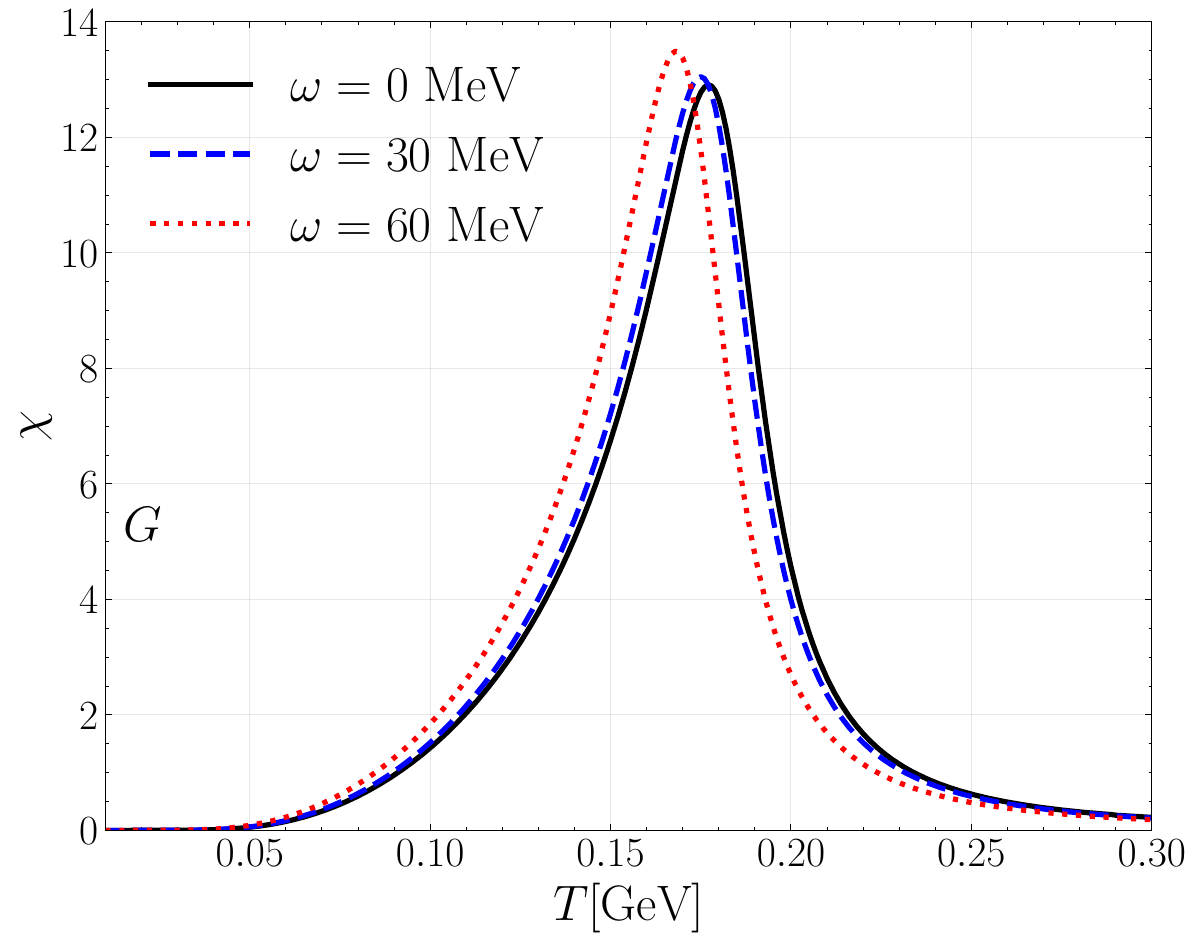}
        \includegraphics[scale= 0.42]{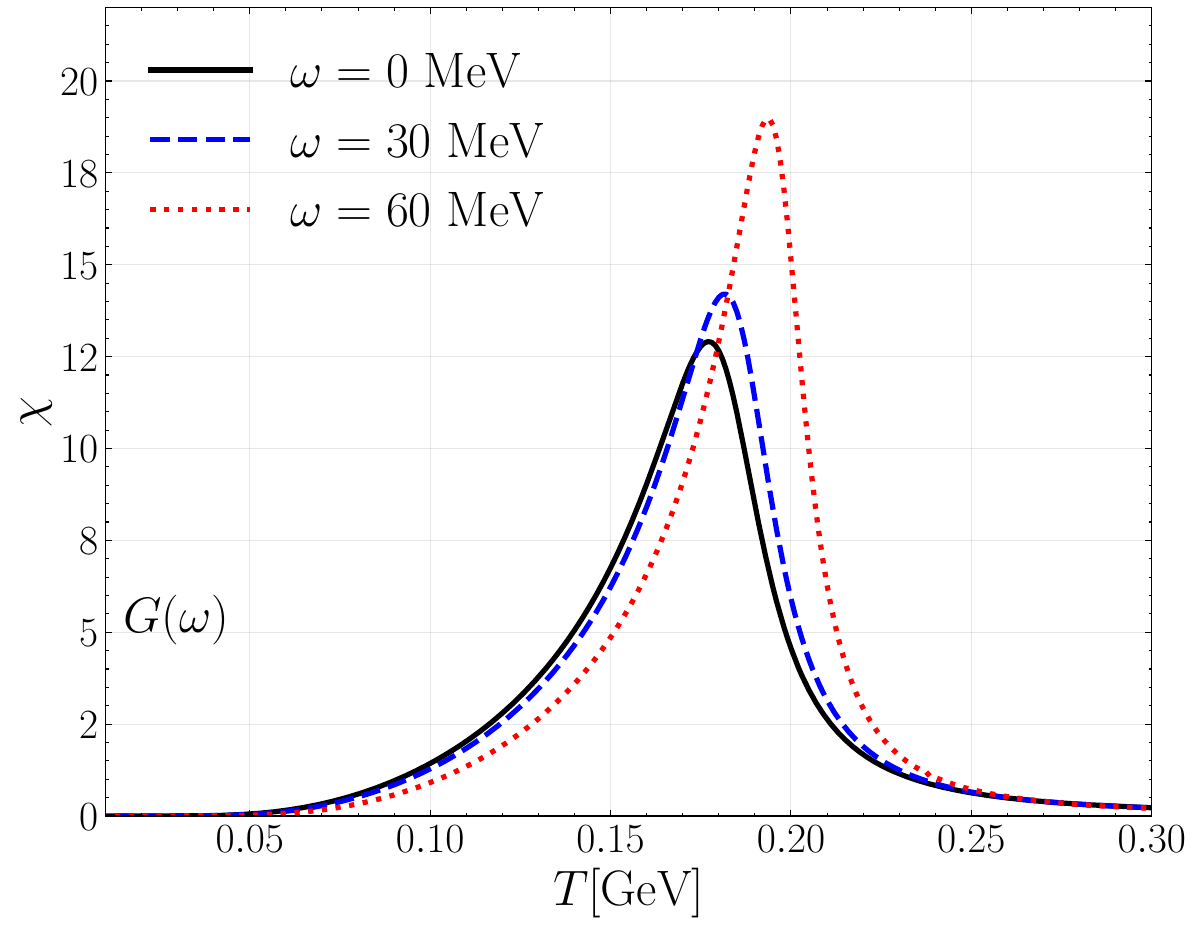}
	\end{center}
	\caption{Chiral susceptibility as a function of the temperature for different values of angular velocity. In the left panel we assume fixed $G$, and in the right panel $G(\omega)$.}\label{dcondxT}
\end{figure}

The behavior of the pseudocritical temperature, $T_{c}$, as a function of angular velocity, $\omega$, at $\mu=0$ is shown in figure \ref{TxW}. Here is clear that the effect of constant and running couplings are opposite, i.e., the pseudocritical temperature decreases with $G$ and increases with $G(\omega)$, which in the former case we observe a qualitative agreement with the PPNJL results shown in Ref. \cite{Sun:2024anu}. In the green band we observe the lattice results, which show that our fitting for $G(\omega)$ indeed gives a very good behavior of the pseudocritical temperature of chiral transition as a function of angular velocity. For comparison, when considering the relation given in Eq.(\ref{TcLattice}), we have from lattice QCD $B_2=1.13153 \pm 0.03$, while for the NJL model we have obtained $B_2 = 1.12813$.


\begin{figure}[!ht]
	\begin{center}
        \includegraphics[scale= 0.42]{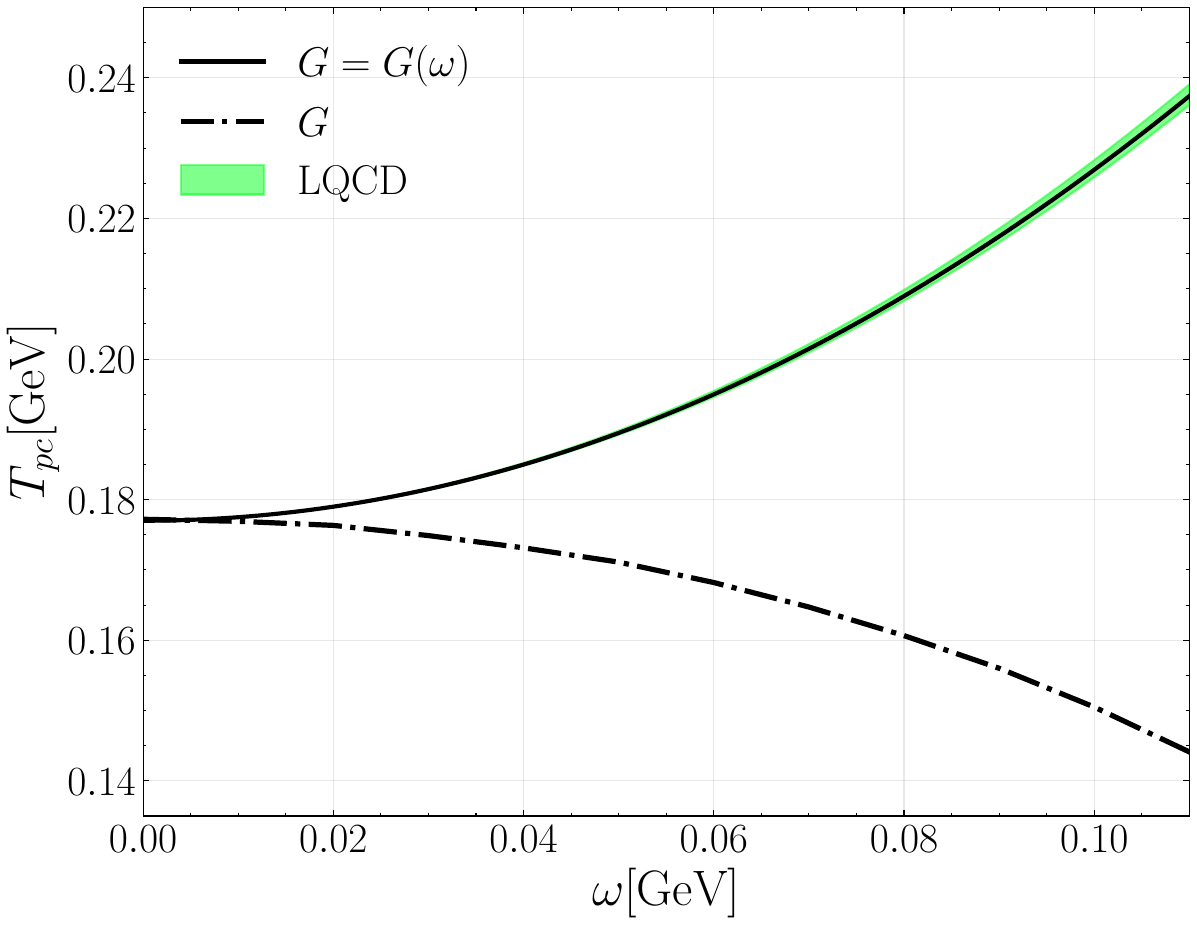}
        \end{center}
	\caption{Pseudocritical temperature of chiral transition as a function of $\omega$ at $\mu=0$, for constant (continuous line) and running (dashed line) couplings. The green band indicate the lattice data transition temperature with its error bar \cite{Braguta:2022str}.}\label{TxW}
\end{figure}


In figure \ref{Txmu} we present the phase diagram $T\times \mu$ with fixed values of $\omega=0, 30
$ and $60$ MeV, for both constant and running couplings.  The transition lines of $(T,\mu)$ with constant coupling decreases its values as we increase the angular velocity. However, when considering the running coupling the transition lines  increases its values with the angular velocity. The CEP appears with a higher value of temperature and a slightly higher value of chemical potential for the running coupling at $\omega=60$ MeV when compared to $\omega=30$ MeV. This result indicates that the effect of rotations with a gluonic effective interaction strengthen the system toward more prominent lines of first-order phase transitions, while there is a shift to lower values of $(T_{CEP},\mu_{CEP})$ when considering the constant coupling. Our phase diagram represents  an expected behavior when we consider the behavior of the transition temperatures in the plane $T\times\omega$ of Figure \ref{TxW}. However, there is some disagreement when considering the findings of NJL model in Ref.\cite{Wang:2018sur}, where
the phase diagram has a expansion of the crossover temperature over the lines of first-order transition as the angular velocity increases. There are recent results for the linear sigma model coupled with quarks, in Ref. \cite{Hernandez:2024nev}, which indicates that the CEP points are shifted to lower chemical potential and higher temperatures as the angular velocity increases, in disagreement with our results. Also, considering the constant coupling, our results show qualitative agreement with the holographic approach from Ref.\cite{Zhao:2022uxc}, where we observe that the transition temperatures and chemical potentials decreases as a function of the angular velocity, while we have a qualitative agreement in the running coupling case with Ref.\cite{Chen:2024jet}.


\begin{figure}[ht!]
	\begin{center}
		\includegraphics[scale= 0.42]{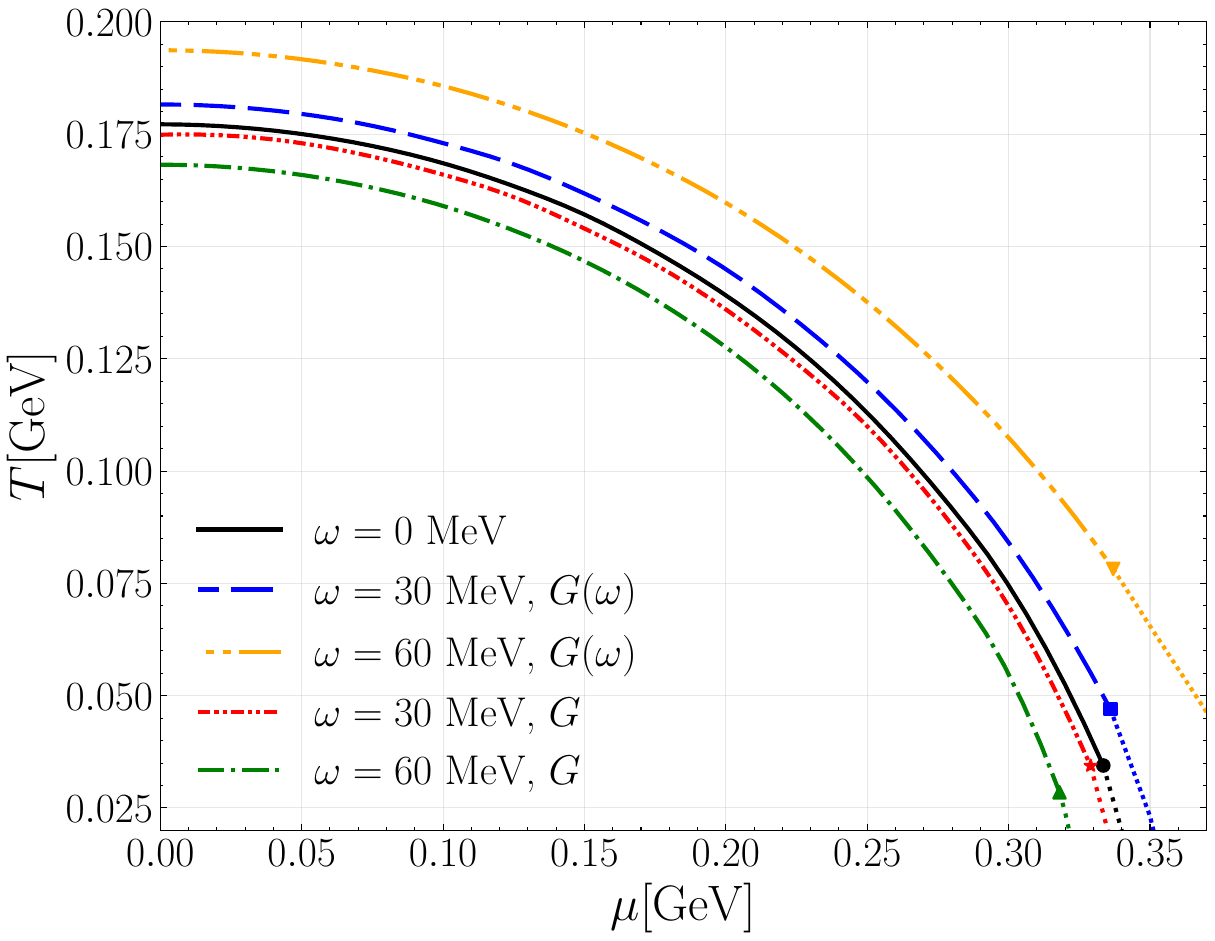}
	\end{center}
	\caption{Phase diagram $Tx\mu$ with fixed $\omega=0,30,60$ MeV, for both constant and running coupling. The big dots are the critical end points (CEPs). After the CEPs, the dotted lines  indicate the first-order region, before the CEPs are the crossover region. }\label{Txmu}
\end{figure}


\section{Conclusion}\label{sec4}

In this work we highlighted the principal aspects of the two-flavor Nambu--Jona-Lasinio model in a rotating framework. 
Recent LQCD data shows that the pseudocritical temperature of chiral and deconfinement transitions have different results when applied to system with quarks and/or gluons: it decreases with quarks; and increases with gluons or with both degrees of freedom. In the former case, the lack of gluons in low energy quark models appears for this problem as unavoidable, inducing us to propose the application of a running coupling as a function of the angular velocity. A seminal application of this idea has been originally studied in Ref. \cite{Jiang:2021izj}, where a coupling with linear dependence on $\omega$ has been suggested based on intuitive semiclassical approach for the QCD coupling with rotation. With the previous approach it is possible to obtain the increasing behavior of $T_c$ as a function of $\omega$. Differently, we match the pseudocritical temperature for chiral transition calculated by LQCD to the NJL model for different values of angular velocity, and then we manage a fitting function for the coupling. This procedure recovers the chiral vortical catalysis, i.e., an enhancement of the chiral condensate by the increasing of the angular velocity, slowing the chiral symmetry restoration by the temperature. However, our fitting function for $G(\omega)$, given by Eq.(\ref{Gw}), enforces the system to reproduce the parabolic behavior for $T_c \times \omega$, observed in LQCD \cite{Braguta:2022str}, which indicates that the linear dependence on $\omega$ just qualitatively captures the gluonic effect in the NJL model as observed in \cite{Jiang:2021izj}. In this way, our results with running coupling indicate a complete opposition with the effects obtained with constant coupling, as well as the NJL results in Ref.\cite{Wang:2018sur} and previous linear sigma model and holographic approaches \cite{Hernandez:2024nev,Chen:2024jet}. Therefore, the chiral symmetry is broken with more intensity as we increasing the angular velocity, a finding which we have not seen in the literature in its full numerical extension, as far as we know. 

The peaks of the chiral susceptibility are an important feature to understand how the fluctuations around the transition behave, and could be important for future detailed analysis. With the constant coupling, the magnitude of the peaks of the susceptibilities changes very little and, as expected, they are shifted to lower temperatures as we increase the angular velocity. On the other hand, with running coupling, the peaks are shifted to higher temperatures
with higher magnitude, indicating stronger fluctuations closer to the crossover. 

The phase diagram has different behaviors with and without the effect of rotations. With the constant coupling, we observe a slow decreasing behavior of the transition lines of $(\mu,T)$ as we increase the angular velocity, while we have an enhancement of the transition temperature with an increasing of the chemical potential, when considering the running coupling.
This last result presents as an interesting prediction, considering that does not agree with recent results from holographic approaches and linear sigma model with quarks.

There is some recent literature discussing the splitting of the pseudocritical temperature of deconfinement and chiral phase transitions, e.g., see Refs.  \cite{Sun:2024anu,Sun:2023kuu,Chernodub:2024wis}, with different approaches and purposes that deviates from the present manuscript. We hope in the near future to report advances on these topics.

\section*{Acknowledgments}

This work was partially supported by Conselho Nacional de Desenvolvimento Cient\'ifico 
e Tecno\-l\'o\-gico  (CNPq), Grants No. 309598/2020-6 (R.L.S.F.); Coordena\c c\~{a}o  de 
Aperfei\c coamento de Pessoal de  N\'{\i}vel Superior - (CAPES) Finance  Code  001 (R.M.N); Fundação Carlos Chagas Filho de Amparo à Pesquisa do 
Estado do Rio de Janeiro (FAPERJ), Grant No.SEI-260003/019544/2022 (W.R.T);
Funda\c{c}\~ao de Amparo \`a Pesquisa do Estado do Rio 
Grande do Sul (FAPERGS), Grants Nos. 19/2551- 0000690-0 and 19/2551-0001948-3 (R.L.S.F.); The work is also part of the project Instituto Nacional de Ci\^encia 
e Tecnologia - F\'isica Nuclear e Aplica\c{c}\~oes (INCT - FNA), Grant No. 464898/2014-5. 
VST is supported by Conselho Nacional de Desenvolvimento Cient\'\i fico e Tecnológico (CNPq), grant 305004/2022-0, and by Funda\c c\~ao de Amparo \`a Pesquisa do Estado de S\~ao Paulo (FAPESP), grant 2019/010889-1. R.L.S.F. is also grateful to A. Roenko and A.Y. Kotov for comments and discussions about the lattice QCD data.

\bibliography{ref.bib}

\end{document}